\documentclass[preprint,12pt]{elsarticle}

\usepackage{epsfig}

\journal{Physics Letters A}

\begin{document}

\begin{frontmatter}

\title{Stationary and dynamical properties of one-dimensional quantum droplets}

\author{Sherzod R. Otajonov}
\author{Eduard N. Tsoy\corref{cor1}}
\author{Fatkhulla Kh. Abdullaev}

\address{Physical-Technical Institute of the Uzbek Academy of Sciences,\\
Chingiz Aytmatov str. 2-B, Tashkent, 100084, Uzbekistan}

\cortext[cor1]{Corresponding author, {\em E-mail}: etsoy@uzsci.net}

\begin{abstract}
  The dynamics of quantum droplets in 1D is analyzed on the basis of the
variational approach (VA). It is shown that the VA based on the
super-Gaussian function gives a good approximation of stationary states. The
period of small oscillations of the perturbed droplet is obtained. It is
found numerically that oscillations are almost undamped for many periods.
Based on the VA, an existence of stable localized states for different
combinations of signs of nonlinearities is demonstrated.
\end{abstract}


\end{frontmatter}

\section{Introduction}
\label{sec:intro}

   There is a considerable progress in theoretical and experimental studies
of quantum droplets (QDs) in ultracold
gases~\cite{Petr15,Petr16,dipexp,Edler,2compexp,Zin,Chei18,Astr18,Liu19,Tononi}.
The existence of the quantum droplets was predicted in Ref.~\cite{Petr15}. It
is well known that a multi-dimensional Bose-Einstein condensate (BEC) with
attraction between atoms experiences a collapse~\cite{Peth11}. Quantum
fluctuations can arrest the collapse in two and three dimensions. Such a
possibility follows from the expression of a correction, known as the
Lee-Huang-Yang (LHY) term~\cite{LHY}, to the ground state energy of Bose gas
due to quantum fluctuations. This correction corresponds to the appearance of
the effective repulsion $\sim n^{5/2}$ in 3D, where $n$ is the condensate
density. The balance between attraction of atoms and the effective repulsion
induced by quantum fluctuations results in the possibility of the quantum
droplet existence. The systems, where quantum droplets are predicted to
present time, include two component BECs~\cite{Petr15,Petr16,Astr18,Liu19}, dipolar
BECs~\cite{dipexp}, Bose-Fermi mixtures, and Bose mixtures with spin-orbit
coupling~\cite{Tononi}. Experimentally QDs are observed in dipolar
BECs~\cite{dipexp,Edler} and Bose-mixtures~\cite{2compexp,Chei18}.

   A study of quasi-one-dimensional BECs represents a particular
interest~\cite{Petr16,Astr18} The form of the LHY term in this case has
nontrivial behavior. While in 2D and 3D LHY term describes the effective
repulsion, for low densities in 1D it corresponds to the effective
attraction. With an increase of the condensate density, the sign of the LHY
term is changed from the attraction to the repulsion~\cite{Edler,Zin}. Thus,
there is a region where localized states in BECs with effective repulsion
between atoms exist due to the effect of quantum fluctuations. A theoretical
analysis of this 1D QDs was performed in recent works~\cite{Petr16,Astr18}.

  In Ref.~\cite{Petr16}, an exact solution for 1D QDs was obtained. In
experiments, an initial distribution of a BEC, formed by external traps, may
not coincide with the exact solution. Therefore, it is important to develop
an approach that describes the dynamics of the QD parameters in time, also in
a presence of different perturbations. We mention that the variational
approach (VA), based on the Gaussian trial function~\cite{Astr18}, works only
for low number of atoms. The purpose of this work is to describe analytically
the dynamical properties of 1D droplets using the modified  variational
approach, based on the super-Gaussian function. We demonstrate that such an
approach gives an excellent description of the QD dynamics. Also, based on
this approach we predict an existence of localized waves (solitons) for a
general case of signs of quadratic and cubic nonlinearities.

   The paper is organized as follows. In Section~\ref{sec:model}, the
modified VA is developed. In particular, the dynamical equations for the QD
parameters are derived, the parameters of stationary QDs are analyzed, and
the frequency of small oscillations of the QD shape is obtained. In
Section~\ref{sec:cases}, an application of the VA to a general case of the
system is presented. Section~\ref{sec:conc} concludes the paper.

\section{The model and dynamical equations}
\label{sec:model}

    Let us consider a two-component BEC under the action of quantum
fluctuations. This system in 1D geometry is described by the following
equations (c.f. Ref.~\cite{Petr16}):
\begin{eqnarray}
   && i \hbar \psi_{1,t} + {\hbar^2 \over 2m} \psi_{1,xx} + (\Gamma_s |\psi_{1}|^2
     + \Gamma_{c} |\psi_{2}|^2) \psi_{1}
\nonumber \\
    &&+ \Delta ( |\psi_{1}|^2 + |\psi_{2}|^2)^{1/2} \psi_{1} = 0,
\nonumber \\
  && i \hbar \psi_{2,t} + {\hbar^2 \over 2m} \psi_{2,xx} + (\Gamma_{c} |\psi_{1}|^2
    + \Gamma_s |\psi_{2}|^2) \psi_{2}
\nonumber \\
    &&+ \Delta (|\psi_{1}|^2 + |\psi_{2}|^2)^{1/2} \psi_{2} = 0,
\label{2comp}
\end{eqnarray}
where $\psi_{1}$ ($\psi_{2}$) is the wave function of the first (second)
component, $\Gamma_s = (3 g + g_c)/2$, $\Gamma_{c} = (g_c - g)/2$ (the self-
and cross-interaction coefficients), and $\Delta = \sqrt{m}\, g^{3/2} /(\pi
\hbar)$ are related to the intra- and inter-species coupling
constants, $g\equiv g_{\uparrow \uparrow} = g_{\downarrow \downarrow} = {2
\hbar^2 a_s /(m a_{\bot}^2) }$ and $g_c \equiv g_{\uparrow \downarrow}$,
respectively, where $g_{\uparrow \downarrow}$ is found similarly to
$g_{\uparrow \uparrow}$ ($g_{\downarrow \downarrow}$) with the corresponding
value of $a_s$. Considering the symmetric case, when $\psi_1 = \psi_2 =
\psi_s \Psi$, the dynamics of the condensate mixture is reduced to the single
Gross-Pitaevskii equation~\cite{Petr16}:
\begin{equation}
  {i \Psi_t} + {1 \over 2 }\, \Psi_{xx} +
  \gamma |\Psi|^2 \Psi + \delta |\Psi| \Psi = 0,
\label{gpe}
\end{equation}
where $\Psi(x,t)$ is the BEC wave function, such that $|\Psi(x,t)|^2$ is the
normalized density in each component. The following scales are used in the
dimensionless Eq.~(\ref{gpe}):
\begin{eqnarray}
  x_s &=& {\hbar \delta \over \Delta} \sqrt{\Gamma_s + \Gamma_c \over 2 m \gamma }, \quad
  t_s = { \hbar (\Gamma_s +\Gamma_c) \delta^2 \over 2 \gamma \Delta^2}.
\nonumber \\
  \psi_s &=& {\sqrt{2} \gamma \Delta  \over (\Gamma_s +\Gamma_c) \delta },
\label{pars}
\end{eqnarray}
where $\gamma$ and $\delta$ are arbitrary constants with $\mathrm{sign}(\gamma) =
\mathrm{sign} (\Gamma_s +\Gamma_c)$ and $\mathrm{sign}(\delta) =
\mathrm{sign} (\Delta)$.

   It is possible to normalize Eq.~(\ref{gpe}) such that coefficients
$|\gamma| = |\delta| = 1$. We prefer to retain a more general notation, since
a case with arbitrary signs of the coefficients will be considered in
Sec.~\ref{sec:cases}. In Refs.~\cite{Petr16,Astr18}, it was shown that a
repulsive condensate $\gamma<0$ with fluctuations $\delta >0$ supports an
ultra-dilute liquid state (quantum droplets) in 1D. In this section, we use
such signs for $\gamma$ and $\delta$.

  Equation~(\ref{gpe}) has an exact solution~\cite{Petr16} that corresponds
to a stationary QD:
\begin{eqnarray}
  \Psi_{\mathrm{ex}} = {-3 \mu \exp ( {-i \mu t} )
    \over \delta + \sqrt{\delta^2 - 9 \gamma \mu/2}\,
    \cosh \! \left(\sqrt{-2\mu}\, x \right) },
\label{ex}
\end{eqnarray}
where $\mu<0$ is the chemical potential. In Ref.~\cite{Petr16}, the solution
is presented for  $\gamma = -1$ and $\delta = 1$. We include an explicit
dependence on $\gamma$ and $\delta$ that will be useful in a discussion of
the general case, see Sect.~\ref{sec:cases}.

   It is worth to note an interesting interpretation of solution~(\ref{ex}).
By using the following relation
\begin{equation}
  \tanh(z + a) +   \tanh(a - z) = {2 \tanh(2a) \over 1 + \mathrm{sech}(2a) \cosh(2z)},
\label{tanh2}
\end{equation}
the droplet solution~(\ref{ex}) can be represented as a combination of two
kinks (a kink and an anti-kink). When $|\mu|$ is small, kinks are close to
each other, and the droplet has a bell shape. In contrast, when $\mu \to
2\delta^2/(9\gamma)$, the kinks are well separated, so the droplet has a flat-top profile.
When $\mu = 0$, we have an annihilation of two kinks.

   As discussed in Sec.~\ref{sec:intro}, we assume that an initial profile of
the BEC density is different from the exact solution. Therefore, one can
expect oscillations and adjustment of the density distribution to the
stationary form. We use the averaged Lagrangian approach~\cite{Ande83} to
find the dynamical equations for the BEC parameters.

  The Lagrangian density for Eq.~(\ref{gpe}) is written as
\begin{equation}
   \mathcal{L} = i (\Psi \Psi_t^{*} - \Psi^{*} \Psi_t) + {1 \over 2} |\Psi_x|^2 -
     {\gamma \over 2}  |\Psi|^4 - {2 \delta \over 3} |\Psi|^3,
\label{lagr}
\end{equation}
where a star means the complex conjugation. We approximate the density
distribution by the super-Gaussian:
\begin{equation}
  \Psi(x,t) = A \exp \left[ - {1 \over 2} \left( {x \over w}\right)^{2m} +
     i b x^2 + i \phi  \right] .
\label{sgauss}
\end{equation}
We assume that QD amplitude $A$, width $w$, chirp $b$, and phase $\phi$ are
dynamical variables, while $m > 0$ is a given parameter that is defined
later. Since $m$ can be fractional, the argument of the super-Gaussian should
be considered as $x^{2m} = (x^2)^m$. The super-Gaussian trial function in
application to the nonlinear Schr\"{o}dinger (NLS) equation was used in a number of
works~\cite{Karl92,Tsoy06,Baiz11}. An inclusion of an additional parameter ($m$)
increases a flexibility of the trial function and accuracy of the
perturbation analysis. At the same time, analytical results become more
cumbersome for interpretation.

  In Eq.~(\ref{gpe}), the total number of particles $N$ is conserved. In
terms of parameters of the trial function~(\ref{sgauss}), $N$ is expressed as
the following:
\begin{equation}
   N \equiv \int_{-\infty}^{\infty} |\Psi|^2 dx = 2\Gamma(1+s) A^2 w,
\label{norm}
\end{equation}
where $s = 1/(2m)$ and $\Gamma(z)$ is the Gamma function. Then the averaged
Lagrangian is found as
\begin{eqnarray}
 L &\equiv& \int_{-\infty}^{\infty} \!\!\! \mathcal{L}\, dx =
   \frac{N\, \Gamma(2-s)}{8 s^2\, \Gamma(s)\, w^2}
   - \frac{\gamma  N^2}{2^{2+s}\, \Gamma(s+1)\, w}
\nonumber \\
   &-& \frac{ 2^{s+1/2}\delta N^{3/2}}{3^{s+1} [\Gamma(s+1)\, w]^{1/2}}
   + \frac{N\, \Gamma(3s)}{\Gamma(s)}\, w^2 (2 b^2 + b_t)
\nonumber \\
   &+& N \phi_t.
\label{avlagr}
\end{eqnarray}
In derivation of Eq.~(\ref{avlagr}), we use expression~(\ref{norm}) for the
norm in order to eliminate $A$. Lagrangian $L$ contains factor
$\Gamma[2-1/(2m)]$ that should be positive. This requirement results in an
additional condition for $m$, namely $m > 1/4$. A variation (the
Euler-Lagrange equations) of $L$ on $w$ and $b$, gives equations for
derivatives $b_t$ and $w_t$, respectively:
\begin{eqnarray}
   && b_t = - {1 \over 8 \Gamma(1+3s)}  \left[ {-3 \Gamma(2-s) \over s w^4 } +
     {3 \gamma \over  2^{s}}  {N \over w^3} \right.
\nonumber \\
     && \left.
     + {2^{3/2 + s} \delta \sqrt{ N \Gamma(s+1)} \over 3^{s}  w^{5/2}}
     + 16 \Gamma(3s+1)\, b^2  \right] \equiv f_b,
\nonumber \\
   && w_t = 2 w b \equiv f_w,
\label{dyneq}
\end{eqnarray}
A variation of $L$ on $N$ and $m$ gives equations for $\phi_t$ and $m$, respectively.

   We stress that the dynamical Eqs.~(\ref{dyneq}) cannot be used directly
since parameter $m$ is unknown. However, one can find a fixed point, that
corresponds to a stationary QD, from the following equations
\begin{eqnarray}
   f_b(w, m, N) &=& 0,
\nonumber \\
   f_m(w, m, N) &\equiv& \left. {\partial L \over \partial m} \right|_{b=0}= 0,
\label{fb_fm}
\end{eqnarray}
taking $b = 0$. The first of Eqs.~(\ref{fb_fm}) is a cubic equation with
respect to $w^{1/2}$ . For all values of $\gamma <0$ and $\delta>0$
considered, this equation has a single positive root $w_s$ that can be found
analytically. Substituting $w_s$ into the second of Eqs.~(\ref{fb_fm}), one
obtains equation for $m_s$. We find that the dynamical Eqs.~(\ref{dyneq})
with $m = m_s$ describe well the parameter variations. For given $N$, one
finds $m_s$, then functions $w(t)$ and $b(t)$ are obtained from
Eqs.~(\ref{dyneq}), while amplitude $A(t)$ is found from Eqs.~(\ref{norm}).

\begin{figure}[htbp]   
  \centerline{ \includegraphics[width=5.5cm]{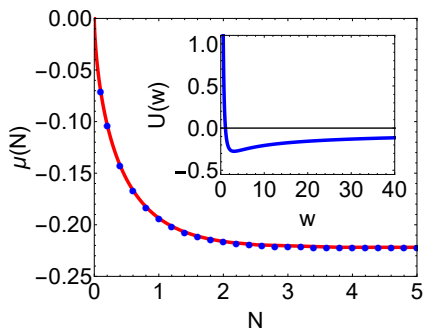}}
\caption{The dependence of the chemical potential $\mu$ on $N$. The line
is found from the exact solution, while points are found from the VA.
The inset shows an effective potential, see Eq.~(\ref{pot}).}
\label{fig:mu}
\end{figure}

   The chemical potential $\mu$ of the stationary solution can be found as
the following
\begin{eqnarray}
   \mu &\equiv& -\phi_{t}(w = w_s) = -{\partial E_{QD} \over \partial N}
     = {-3 \gamma N \over 2^{s+3} \Gamma(s+1)\, w_s}
\nonumber \\
     &&  - {5\cdot 2^{s-1} \delta N^{1/2} \over 3^{s+1} [2 \Gamma(s+1) w_s]^{1/2}},
\label{mu}
\end{eqnarray}
where $E_{QD} = N \phi_t - L$ at $w = w_s$ and $b = 0$ is the energy of the
stationary QD. The dependence of $\mu$ on $N$ is presented in
Fig.~\ref{fig:mu}. The line represents $\mu(N)$ found from the exact
solution~(\ref{ex}), and points show the results of the VA. Since $d \mu / d
N < 0$, the stationary QD is stable according to the Vakhitov-Kolokolov
criterion~\cite{Vakh73}.

\begin{figure}[htbp]   
\centerline{ \includegraphics[width=5.5cm]{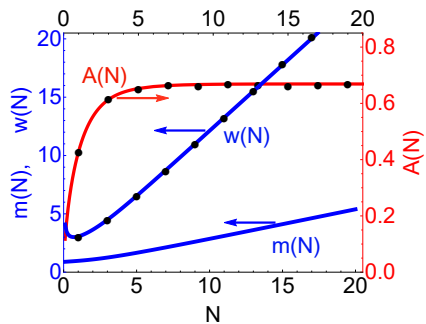}}
\caption{Dependencies of droplet width $w$, the power exponent $m$ (the left
axis), and the droplet amplitude $A$ (the right axis) on $N$ for $\gamma =
-1$ and $\delta = 1$. Lines correspond to the VA, while points are from
numerical simulations of Eq.~(\ref{gpe}). } \label{fig:pars}
\end{figure}

   In Fig.~\ref{fig:pars}, we compare parameters, found from the VA (lines),
with those, obtained from direct numerical simulations of Eq.~(\ref{gpe})
(points). In simulations, we employ Eq.~(\ref{sgauss}) as an initial
condition. For given $N$, we find $m_0=m_s$ and parameters of the stationary QD
$w_s$ and $A_s$. We take initial parameters as $w_0 = 1.1 w_s$ $A_0 =
A_s/\sqrt{1.1}$ so that the initial norm does not change. This initial
condition results in almost periodic variations of the width and amplitude.
After some period of time ($t \sim$ 500-1000), we measure the average of the
maximum and minimum amplitudes over the period. A similar procedure is
applied for finding the QD width.

   One can see that the droplet amplitude Fig.~\ref{fig:pars} tends to 2/3
for large $N$, as follows from the exact solution~(\ref{ex}). The droplet
width increases almost linearly at large $N$. Therefore, QDs demonstrate the
incompressibility similar to ordinary liquids. For $N \to 0$, $m \to 0.877$.
The shape of QDs is close to the Gaussian ($m \approx 1$) for $N \approx
1.5$. For $N <1.5$, the QD shape has a sharp peak, while for $N > 1.5$, the
shape tends to a flat-top profile. Figures~\ref{fig:mu} and~\ref{fig:pars}
show that the VA based on the super-Gaussian gives excellent results for
parameters of stationary droplets. In contrast, the VA based on the Gaussian
function [a fixed $m$ = 1 in Eq.~(\ref{sgauss})] gives a poor
approximation~\cite{Astr18}  of stationary solutions, valid only for $N \ll
1$.

  By eliminating $b$ from Eqs.~(\ref{dyneq}), equation for $w(t)$ is
represented as an equation of motion of an effective particle in a potential:
\begin{equation}
   w_{tt} = - {\partial U(w) \over \partial w},
\label{w_tt}
\end{equation}
where
\begin{eqnarray}
   U(w) =
     {1 \over \Gamma(3s)} \left[
     {\Gamma(2-s) \over 8 s^2 w^2} - {\gamma N \over 2^{s+2} s w}
     \right.
\nonumber \\
     \left.
     - {2^{s+1} \delta \over 3^{s+1}}   \left({N\, \Gamma(s)\over 2 s w } \right)^{1/2}
     \right].
\label{pot}
\end{eqnarray}
The potential $U(w)$ for  $N= 1$ and $m= 0.971$ is shown in the inset of
Fig.~(\ref{fig:mu}). For given $N$ and $m$, potential $U(w)$ has a single
minimum at $w = w_s$ that corresponds to the stationary QD. Potential $U(w)$
tends to zero at large $w$. The VA also describes the dynamics of the QD
parameters, when the initial profile is different from the stationary one.

   When the initial energy $\mathcal{E}_0= w_t^2(0)/2 + U(w_0)$ of the
effective particle is negative, the VA predicts a periodic variation of the
QD width (and other parameters), see Eq.~(\ref{w_tt}). However, this picture
is true only for small deviations from the stationary QD shape. For moderate
deviations ($>$ 20\%-30\%), numerical simulations of Eq.~(\ref{gpe}) show a
generation of linear waves and a splitting of the initial distribution. This
means that a moderately deformed QD breaks into several droplets. Some of
these droplets move in opposite directions with the same velocities, such
that the total momentum is conserved.

\begin{figure}[htbp]   
\centerline{\includegraphics[width=6.cm]{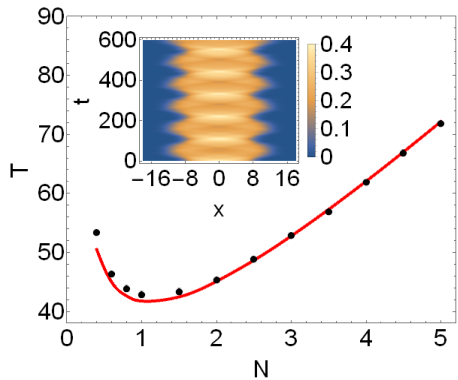}} \caption{The period of
small oscillations $T$ of the QD width, found from Eq.~(\ref{omega}) (the
line) and from numerical simulations (points) of Eq.~(\ref{gpe}), as a
function of $N$ for $\gamma = -1$ and $\delta = 1$. The inset shows a map
plot of $|\Psi(x,t)|^2$ for $N = 5$ , $m_s= 1.66$ ($A_s=0.653$ and
$w_s=6.54$). }
\label{fig:per}
\end{figure}

   A variation of the QD shape, found from numerical simulations of
Eq.~(\ref{gpe}), is presented in the inset of Fig.~\ref{fig:per}. Similar to
Fig.~\ref{fig:pars}, we change the parameters of the stationary profile by 10
\%, and use this profile as an initial condition. We monitor the QD dynamics
over $t \sim 800$ and find the period as an average on the last 2-4
oscillations.

   We should emphasize that oscillations of the QD shape near the stationary
shape are almost undamped. We perform numerical simulations of
Eq.~(\ref{gpe}) up to $t \sim 10000$ ($\sim$ 200 periods), and we do not
observe substantial decrease of the oscillation amplitude, after some initial
shape adjustment. A sustainability  of oscillations is a peculiar property of
QDs, and it can be used to distinguish them from the NLS solitons.

   By using the potential, we can find the frequency $\Omega_0$ of small
oscillations of droplet parameters near the stationary state as
\begin{equation}
   \Omega_0^2 = \left. {\partial^2 U(w) \over \partial w^2} \right|_{w= w_s} .
\label{omega}
\end{equation}
Figure~\ref{fig:per} shows the dependence of the period of oscillations
$T= 2\pi/\Omega_0$ found from Eq.~(\ref{omega}) and from
numerical simulations.

\begin{figure}[htbp]   
\centerline{\includegraphics[width=5.5cm]{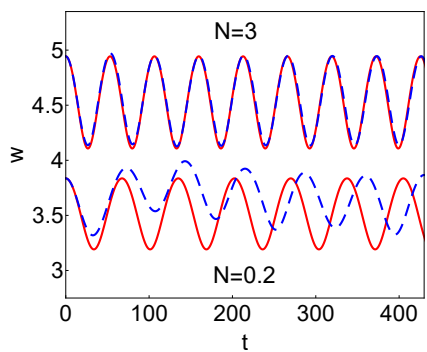}} \caption{Oscillations of
the QD width found from numerical simulations (solid lines) of Eq.~(\ref{gpe})
and from Eqs.~(\ref{w_tt}) of the VA (dashed lines).}
\label{fig:w_t}
\end{figure}

   The dynamics of the QD width, found from numerical simulations of
Eq.~(\ref{gpe}), is compared with predictions of the VA, Eq.~(\ref{w_tt}), in
Fig.~\ref{fig:w_t}. One can see that for small $N$ ($N = 0.2$ in the figure)
there is a noticeable difference between the results, see also
Fig.~\ref{fig:per}. For such $N$, the actual profile is different from
distribution~(\ref{sgauss}). Therefore, an initial adjustment of the QD shape
shifts the phase of oscillations. Nevertheless, there is a reasonable
qualitative agreement of results for the amplitude and the period of
oscillations. For $N > 1$, an agreement between numerical simulations and the
VA is good, as seen from Figs.~\ref{fig:per} and~\ref{fig:w_t}. Results,
summarized in Figs.~\ref{fig:mu}-\ref{fig:w_t}, shows that the VA correctly
predicts not only stationary parameters, but also the dynamics of QDs.

   When $\mathcal{E}_0 > 0$, according to Eq.~(\ref{w_tt}), a pulse is
broadened infinitely, and disappears. This prediction is valid only
qualitatively. Numerical simulations of Eq.~(\ref{gpe}) indicate that the
regime of splitting is replaced by the dispersive broadening for much larger
deformations, than the VA predicts. For deformations that preserve $N$, a
large deformation corresponds to small values of the distribution area $S =
\int_{-\infty}^{\infty} |\psi| dx$. This indicates that there is a threshold
value $S_{th}$ such that when $S_0 < S_{th}$, an initial distribution is
broadened. This property is similar to the property of pulses in the NLS
model, $\gamma = 1$ and $\delta = 0$ in Eq.~(\ref{gpe}), where threshold
value is given by $S_{th,NLS} = \pi/2$.

   Let us estimate parameters for realistic experiments in a binary BEC of Rb
atoms ($m = 1.42\cdot10^{-25}\ \mathrm{kg}$) in spin up and spin down states.
For the intra-species scattering length, we take
$a_{\uparrow\uparrow}=a_{\downarrow\downarrow}= 2000 a_{0}$, and assume that
the inter-species scattering length $a_{\uparrow\downarrow} \sim
-(0.95\mathrm{-}0.99) a_{\uparrow\uparrow}$, where $a_0$ is the Bohr radius,
so that the residual scattering length satisfies $0 < a_{\uparrow\downarrow}
+ \sqrt{a_{\uparrow\uparrow} a_{\downarrow\downarrow}} \ll
\sqrt{a_{\uparrow\uparrow} a_{\downarrow\downarrow}}$~\cite{Petr16}. The
transverse radius of a trap is taken as $a_\bot = 0.6 \ \mu\mathrm{m}$. Then,
the critical temperature of the Bose condensation for a gas with density
$\sim 10^{14}\ \mathrm{cm}^{-3}$ is $T_{cr}= 0.7\ \mu\mathrm{K}$. The
characteristic scales for a system with $|\gamma| =1$ and $|\delta| =1$ are
$x_s \sim (1\mathrm{-}0.4)\ \mu\mathrm{m}$, $t_s \sim (0.8\mathrm{-}0.2)\
\mathrm{ms}$, and $N_s \sim 40\mathrm{-}450$. These parameters are compatible
with typical experiments on BECs~\cite{Peth11}.

  As far as we know, there are no experiments on 1D QDs.  In experiments on
3D QDs in potassium mixtures~\cite{2compexp,Chei18}, a size of QDs is
$0.5\mathrm{-}6\ \mu\mathrm{m}$, and the total number of particles $N \sim
(2\mathrm{-}25)\!\cdot\!10^{3}$. Our estimation shows that it is more easy to
create 1D QDs in BECs with heavier atoms and larger scattering lengths. This
motivates our choice of Rb in calculation of characteristic scales.

\section{Dynamics of solitons for different values of $\gamma$ and $\delta$}
\label{sec:cases}

  Equation~(\ref{w_tt}) for $w$ with potential~(\ref{pot}) allows us to study
the dynamics of localized waves for different values of $\gamma$ and
$\delta$. In BECs, the sign of $\gamma$ can be changed via the Feshbach
resonance, see e.g.~\cite{Peth11}. Though only a positive sign of $\delta$
was considered in Ref.~\cite{Petr16}, here we assume that $\delta$ can have
any sign. By normalization of Eq.~(\ref{gpe}), one can eliminate the
dependence on absolute values of the parameters. Therefore, we consider
cases, when $\gamma$ and $\delta$ each take one of the three values  -1, 0,
and 1. When $\delta = 0$, we either have a free linear condensate ($\gamma
=0$), or a well-studied system of a BEC with the two-body
interaction~\cite{Peth11}. Since a system with $\gamma <0$ and $\delta > 0$
is considered in the previous section, we only need to study five cases. In
BEC literature, QDs, localized states in a presence of quantum fluctuations,
are distinguished from solitons that exist without fluctuations. However, in
this Section, we call all localized states as solitons.

\begin{figure}[htbp]   
  \centerline{ \includegraphics[width=6cm]{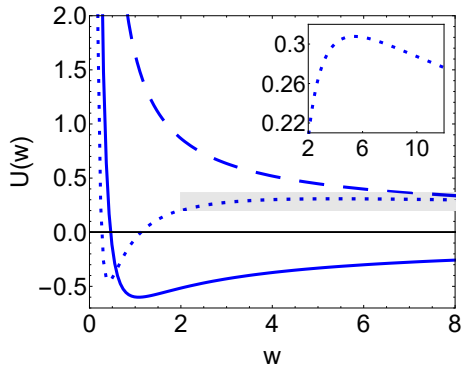}}
\caption{Shapes of potentials for different $\gamma$ and $\delta$. Parameters
($\gamma$, $\delta$, $N$, $m$) for the solid, dashed and dotted lines are (1,
1, 1, 0.842), (-1, -1, 1, 0.971), and (1, -1, 7, 0.782), respectively. The
inset shows a scaled-up region (a grey rectangle) around the maximum of the
dotted curve.
}
\label{fig:3pots}
\end{figure}

  Typical shapes of the potentials are presented in Fig.~\ref{fig:3pots}. All
potentials in Fig.~\ref{fig:3pots} tend to zero at $w \to \infty$. For
$\delta = 1$ and  $\gamma= 0$, or $\gamma= 1$ (the solid line in
Fig.~\ref{fig:3pots}), the shape of $U(w)$ is the same as presented in the
inset of Fig.~\ref{fig:mu}. This means that when $\delta > 0$, there exists a
stable soliton for any values of $\gamma$. However, there are no flat-top
solitons when $\gamma \geq 0$, all solitons are bell-shaped with $m \sim
0.8-0.9$. In particular, for $\gamma= 0$ and $\delta = 1$, the power exponent
$m= 0.877$ does not depend on $N$, c.f. Eq.~(\ref{sech}).

  When $\gamma \leq 0$  and  $\delta < 0$ (the dashed line in
Fig.~\ref{fig:3pots}), there are no bright solitons, therefore any initial
pulse spreads dispersively. This is not surprising, because for these
parameters, both nonlinearities correspond to repulsive self-interaction. For
monotonically decreasing potential $U(w)$, there is a question what number
$m$ should be used in the dynamical equation~(\ref{dyneq}), since there are
no stationary solitons, or proper roots for $m$ in the second of
Eqs.~(\ref{fb_fm}). In this case, in order to model the dynamics, one can use
the initial value $m = m_0$ in Eq.~(\ref{w_tt}).

   For case  $\gamma> 0$ and $\delta < 0$, there are two choices: (i) when $N
< N_{\mathrm{th}}$, no solitons exist (with a form of the potential similar
to the dashed line in Fig.~\ref{fig:3pots}), (ii) when $N \geq
N_{\mathrm{th}}$, two stationary solitons exist, because potential $U(w)$ has
two extrema (see the dotted line and an inset in Fig.~\ref{fig:3pots}). In
fact, for these signs of $\gamma$ and $\delta$, and  $N > N_{\mathrm{th}}$,
the second of Eqs.~(\ref{fb_fm}) has two roots for $m$, however one root
corresponds to unstable solitons with $\mu >0$. The threshold value of the
norm for $\gamma = 1$ and $\delta = -1$ is $N_{\mathrm{th}} \approx 4.2$. As
follows from Eq.~(\ref{pot}), $U(w)$ tends to zero at $w \to \infty$ from
positive values. We also mention that the value of the potential at the
minimum can be negative or positive, depending on $N$.

  Numerical simulations shows that for $\gamma> 0$ and $\delta < 0$ there is
no splitting of initially deformed solitons. A deformed distribution emits
linear waves and adjusts its form to the soliton profile. Also, for large
deformations, there is a broadening of initial distributions.

  We obtain that the exact solution~(\ref{ex}), derived in Ref.~\cite{Petr16}
for $\gamma < 0$ and $\delta > 0$, can be generalized for any signs of
$\gamma$ and $\delta$, provided that the soliton exists. When $\gamma = 0$
and $\delta > 0$, solution~(\ref{ex}) is reduced to~\cite{Liu19,Tononi}
\begin{equation}
   \Psi(x,t) = -{3 \mu \over 2 \delta}\, \mathrm{sech}^2(\sqrt{-\mu/2}\, x) e^{-i\mu t}.
\label{sech}
\end{equation}
It follows from the analysis of the exact solutions~(\ref{ex}) and
(\ref{sech}), and also from the VA that solitons are stable in the whole
region of existence because $d \mu /dN <0$.

  The predictions of the VA are supported by numerical simulations of
Eq.(\ref{gpe}). Therefore, the VA based on the super-Gaussian function
describes well the dynamics of localized states for any values of $\gamma$
and $\delta$.

\section{Conclusions}
\label{sec:conc}

  The variational approach based on the super-Gaussian function has been
developed for a description of the dynamics of quantum droplets. A comparison of
VA predictions with results of numerical simulations shows an excellent
agreement for stationary parameters of QDs. The period of small oscillations
of the soliton shape has been found. The oscillation period of quantum
droplets is much larger than the characteristic time scale. The long-lived
oscillations of the QD shape indicates an existence of a linear mode
localized on the QD.

  It has been demonstrated that the VA provides a correct prediction of
existence of stable localized states for any values of $\gamma$ and $\delta$.
It has been found that for  $\gamma> 0$ and $\delta < 0$, solitons are formed
when the norm exceeds the threshold value. Numerical simulations of
Eq.~(\ref{gpe}) shows a splitting of a moderately deformed QD, and dispersive
broadening for large deformations.

  Good agreement of theoretical results with numerical
simulations show that the super-Gaussian function is close to
the actual shape of localized waves in the system. Therefore,
this trial function can also be used in the analysis of the
droplet dynamics under various perturbations.

\section*{Acknowledgements}
  This work has been supported by grant FA-F2-004 of the
Ministry of Innovative Development of the Republic of Uzbekistan.

\end{document}